# A Poynting-Robertson-like drag at the Sun's surface

Ian Cunnyngham[1], Marcelo Emilio[2], Jeff Kuhn*[1], Isabelle Scholl[1], Rock Bush[3]

[1] Institute for Astronomy, University of Hawaii, 34 Ohia Ku St., Pukalani, Maui, HI, 96790, USA

[2] Ponta Grossa State University, Ponta Grossa Parana, Brazil, 84030-900.

[3] Stanford University, Stanford, CA 94305, USA

*Correspondence to: jeff.reykuhn@yahoo.com

## ABSTRACT

The Sun's internal rotation $\Omega(r,\Theta)$ has previously been measured using helioseismology techniques and found to be a complex function of co-latitude, θ, and radius, r. From helioseismology and observations of apparently "rooted" solar magnetic tracers we know that the surface rotates more slowly than much of the interior. The cause of this slow-down is not understood but it is important for understanding stellar rotation generally and any plausible theory of the solar interior. A new analysis using 5-min solar p-mode limb oscillations as a rotation "tracer" finds an even larger velocity gradient in a thin region at the top of the photosphere. This shear occurs where the solar atmosphere radiates energy and angular momentum. We suggest that the net effect of the photospheric angular momentum loss is similar to Poynting-Robertson "photon braking" on, for example, Sun-orbiting dust. The resultant photospheric torque is readily computed and, over the Sun's lifetime, is found to be comparable to the apparent angular momentum deficit in the near-surface shear layer.

## I. INTRODUCTION

One surprise from helioseismic p-mode frequency inversion studies [1-6] has been the detection of a near-surface rotation gradient in the outer 5% of the convection zone. These studies have coarse radial resolution, >3000km (much larger than the 150km density scale length at the photosphere), and find that the rotation decreases outward as dlogΩ/dlogr = α where $\alpha \approx -1$ near the equator, increasing (in magnitude) toward the poles. In contrast, conservation of angular momentum in overturning convective elements would suggest $\alpha \approx -2$ [5]. Neither this near-surface shear, nor the deeper rotation shear in the tachocline at the base of the convection zone have been fully explained [7-9].

Full-disk spatially resolved observations of the Doppler shift of Fraunhofer absorption lines directly yield the surface rotation [10,11], as does timing the motion of magnetic and non-magnetic features as they rotate across the solar disk [11]. Systematic variations in these rotation rates are often interpreted as evidence of this radial gradient, with seemingly faster-rotating surface features "anchored" deeper in an atmosphere, rotating faster than the outer region sampled by Doppler data [11]. Supergranulation, sunspots, and active regions also exhibit a spread in rotation velocity of a few percent. Whether or not this is evidence of a range in anchor depths is unclear. Doppler measurements of different Fraunhofer lines have not directly seen a surface rotation gradient [10].

We measured individual acoustic oscillations (p-modes) in a narrow annulus around the solar limb. This is possible because the Solar Dynamics Observatory/ Helioseismic Observatory (SDO/HMI) satellite is above the effects of the Earth's atmosphere. Our limb darkening function (LDF) observations can accurately measure solar atmospheric structure because of the unblurred tangential line-of-sight through the extreme solar limb.

## II. LIMB ASTROMETRY AND OSCILLATIONS

From space limb solar oscillations can be measured to microarcsecond positional and $10^{-6}$ relative brightness accuracy with the HMI [12]. This instrument obtains 4K x 4K pixel full-disk images in narrow wavelength passbands over a range of linear and circular optical polarization states every 45 seconds [13]. We use six 7.6 pm-wide passbands that are spaced in wavelength by 7.0 pm steps across an Iron Fraunhofer line at a central wavelength of 617.334 nm. We analyzed several polarization states in each of the six filtergram timeseries to obtain 12 independent measurements of the limb brightness $\alpha_i(\theta,t)$ and position, $\beta_i(\theta,t)$ *(i=0...11)* following the techniques described in ref. *12*. For these measurements the different polarization states provide independent datasets with no apparent polarization dependence. The mean limb position (solar radius) is derived from the average of β$_i$ over θ. Fig. 1 shows the apparent solar radius in each filter over the 3.5 year duration of these data. This limb displacement is largest (biggest solar radius) at the line core and varies with observation time because the solar-frame filter wavelengths vary with the orbital Doppler shift of the satellite. This radius variation is also consistent with the apparent radius variations derived from HMI Venus transit timing data using a different analysis [14]. We compute the effective height of each dataset, *i*, from the Fig. 1 data. Each samples a, partially overlapping, vertical range in the atmosphere of about 50km due to the satellite's range of orbital velocity with time and the spread in formation height of the absorption line. *Disk center* models of the theoretical formation of the Iron line at its central wavelength give a mean formation height of about 250km above the atmospheric reference level where the continuum opacity is unity [15,16]. The temperature minimum in these solar atmosphere models occurs at a height of about 500km.

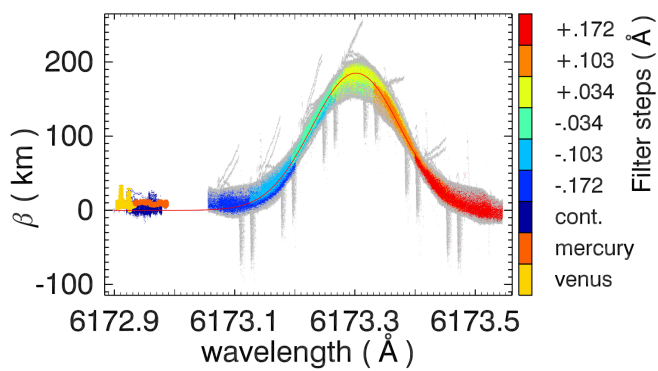

Fig 1: Measured solar radius variation versus filter wavelength during a 3.5 years duration observing window. Colors indicate HMI filtergrams and their nominal central wavelength steps in Angstroms are indicated on the right axis. The HMI continuum data are indicated with their observing campaign labels 'mercury' and 'venus.'

Solar p-mode oscillations generate brightness $\alpha(\theta,t)$, and shape, $\beta(\theta,t)$, perturbations at the limb. As this surface structure rotates over the limb onto or off the visible solar disk the projection in the plane of the sky causes this oscillating limb structure to appear to rotate clockwise or counterclockwise along the limb at a rate $\Psi$, depending on whether the north rotation axis is pointed into or out of the plane. We use the fact that the Sun's axis is inclined by $B=7.2$ degrees from the normal to the ecliptic to determine the mean solar rotation at the atmospheric depth of the limb structure.

From the Earth the Sun's axis appears to precess around the ecliptic normal direction with a one-year period, while it moves in and out of the plane of the sky by an angle $\gamma$ determined by $sin(\gamma) = sin(B)sin(2\pi t/P)$. For small angles $B$ and $\gamma$ this geometric temporal modulation varies as $\gamma(t) = Bsin(2\pi t/P)$ where $P$ is the Earth's synodic orbit period, and $t$ is time. The rate at which structure appears to rotate along the limb is then given by

$$\Psi(t) = \Omega(r)sin(\gamma(t)) \quad (1)$$

where the mean (latitudinal average) solar rotation rate at radius $r$ is $\Omega(r)$.

The oscillating structure is spatially isotropic in the local solar frame. Our analysis Fourier analyzes the timeseries into 'limb" angular and temporal frequency harmonics of the form $\alpha(\theta,t) = Real(\sum_{k,\omega} \tilde{\alpha}(k,\omega) \exp(i(k\theta - \omega t)))$. The apparent rotation along the limb perturbs the p-mode temporal frequencies in proportion to $k$ by

$$\delta\omega_k = 2\pi\delta\nu_k = \Psi(t)k = \Omega(r)\ sin(B)\ sin\left(\frac{2\pi t}{P}\right)k. \quad (2)$$

Twelve, 3.5 year duration, 45s cadence timeseries yield k-ω power spectra, $\left|\widetilde{\alpha_\iota(k,\omega_s)}\right|^2$ that reveal individual p-mode frequencies. Limb harmonics from the 256 angle bins are indexed by $k=0...255$ and temporal frequency by $\omega_s = \frac{2\pi s}{T}$. There are a total of about $2\times10^6$ time-domain points in each set.

Acoustic p-modes with 5-min periods brighten and displace the photosphere with cyclic frequencies $\nu_{nlm} = \omega_{nlm}/2\pi$ where $n$, $l$, and $m$ are the radial, angular and azimuthal spherical harmonic mode indices. For example, the solar surface displacement due to an (*nlm*) p-mode has the spatial form $\delta r(\theta,\phi,t) = Real(a_{nlm}Y_{lm}(\theta,\phi)\exp(i\omega_{nlm}t))$ with $Y_{lm}$ a spherical harmonic [17]. The mode displacement amplitudes are of order 10 microarcseconds, which corresponds to modal relative brightness amplitudes of about $10^{-6}$. Since these HMI data only sample the oscillations along the limb, it can be shown that each spherical harmonic contributes to the limb oscillation power over a range $k \leq l$ in these spectra. In general this "overlap" between spherical harmonics and limb harmonics is greatest for $k=l$ and $m=0$. Fig. 2 shows an example limb k-ν power spectrum. Full-disk p-mode frequencies for *l=k* and *m=0* are over-plotted to show the agreement with conventional Doppler helioseismology [18]. Each ridge structure that runs from the lower left to upper right here corresponds to a fixed radial p-mode node number, *n*. Only the highest amplitude, and highest *k=l* data point in each "parabolic" power distribution corresponds to the frequency of the global p-mode.

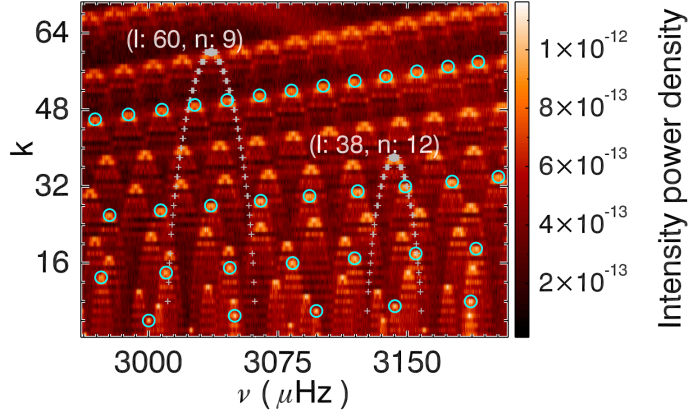

Fig. 2: Limb power spectrum, $|\tilde{\alpha}(k,\nu)|^2$, for line core filter data. Overplotted open circles show observed global m=0 mode frequencies [18]. Blue cross symbols show selected mode leakage calculations of some $m \neq 0$ full-disk modes for the indicated angular ($l$) and radial ($n$) modes into the limb harmonics, $k$, indicated on the vertical axis. The color scale legend is indicated on the right in units of fractional brightness fluctuation (squared) per frequency bin.

## III. ROTATION AND OSCILLATIONS

To derive $\Psi_i(t)$ from Eq. (2) for each filter passband, $i$, we computed spectra in running data blocks of 90 days duration, and fit the observed time-dependent frequency changes of the peaks in all constant-n radial ridges as in Fig. 2. We then use Eq. (2) to fit the power spectra peak frequencies, $\delta\omega(t)$, to obtain the limb rotation $\Psi_i(t)$ versus time. A correction for the sliding boxcar average over a sinusoid is also applied to recover the infinite resolution sinusoid amplitude. Figure 3 shows how the derived limb rotation rate $\Psi_i(t)$ from the linecenter data agree with the sinusoidal modulation of the Sun's rotation axis projection. Each dataset samples a different height range within the photosphere allowing a measurement of the near-surface rotation shear.

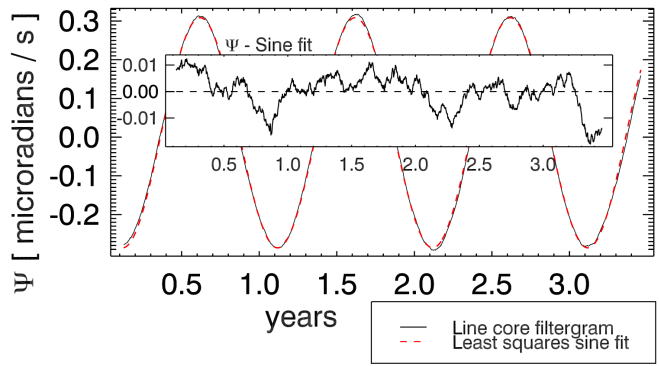

Fig. 3: Example sinusoid fit to limb seasonal rotation $\Psi_i(t)$ temporal variation for the line core filter data. The inset graph shows the residual between the data and sinusoidal rotation model

The circular average mean solar rotation is derived from the amplitude of the temporal sinusoidal variation of $\Psi_i(t)$ from Eq. (1). The total vertical range of all the limb measurements is about 120 km in the photosphere and the 12 measurements with six filter wavelengths and several polarization states cluster into three different heights. Table 1 shows measured height with respect to the continuum, and the rotation rate and its standard error determined from the spread of each set of 4 measurements. Depth and latitude smearing will spatially average the local rotation gradient. Higher resolution could detect an even larger rotation gradient. Fig. 4 shows our limb rotation results compared to Doppler and GONG full-disk p-mode inversion rotation results [4] on logarithmic vertical and horizontal scales. This photospheric shear is larger than has been measured anywhere in the interior. Our mean rotation is slightly slower than the Doppler data, but it is also difficult to localize the effective height of those data.

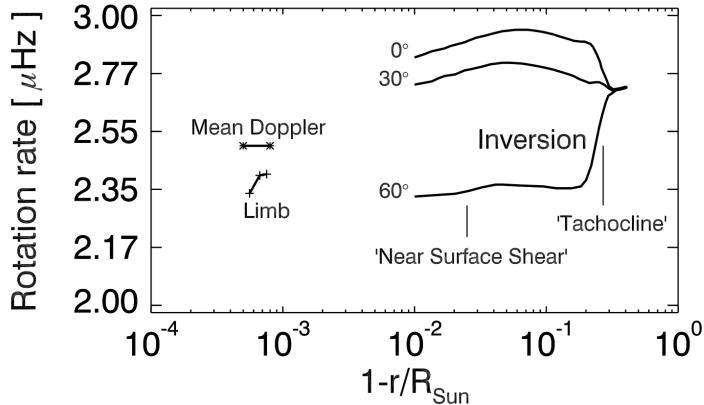

Fig. 4, Solar angular rotation rate versus depth and colatitude on a log-log scale derived from ref. 4. The final inverted near-surface rotation from the two major helioseismology experiments, refs. 3 and 4, are consistent. 'Limb' points (cross symbols) show our observed mean solar rotation shear; 'Mean Doppler' (star symbols) indicate the angle-averaged Doppler rotation over the indicated indeterminate depth range; 'Inversion' shows the p-mode inversion rotation rates at three latitudes through the interior, with the near-surface and tachocline shear zones annotated. The Sun's temperature minimum defines the outer reference radius (Rsun) for the horizontal scale.

## IV.    DISCUSSION AND CONCLUSIONS

It is interesting that we find this large velocity shear in a radiative region where the solar atmosphere is becoming transparent. Here the rotating photospheric plasma radiates photons from below without recapturing their momenta and energy. Consider the total angular momentum contained within the Sun's diffusive photon distribution. The surface photospheric radiation continually diminishes this (Eq. 3) because of a steady outward angular momentum current. Since the local plasma rotation velocity within the Sun is the same for the matter and photons, this coupling implies that the radiated angular momentum must diminish the plasma angular momentum.

The angular momentum reservoir increases rapidly with depth because of the Sun's exponential density stratification. Correspondingly, the outward angular momentum current implies a rapidly decreasing velocity perturbation toward the interior. It follows that the radiated angular momentum has its largest effect on the rotation velocity near the surface. The rotation drag can hardly affect the local vertical stratification of the atmosphere because this is determined at lowest order by the gravitational potential and the outward energy flux.

The notion that an external torque determines the mean solar rotation is not new. Kraft [19] convincingly argued that the Sun lost most of its angular momentum during its lifetime because of the external magnetic braking torque of the solar wind. Speculation that the present differential rotation is, in part, a relic of a latitudinal dependence to this torque is countered by the common view that the current angular rotation is rapidly determined by the *internal* non-diagonal Reynolds stresses created by the interaction of convection and rotation in a highly stratified stellar envelop [7-9,20,21]. While none of these models yet reproduces the solar rotation data in Fig. 4, there are theoretical and anelastic numerical models that generate, for example, an equatorial acceleration [20,21] and tachocline shear [21]. In the spirit of "mean field" theories that average over the effects of convective eddies we suggest here that even a weak radiative torque acting at the photosphere, when integrated over the solar lifetime, can account for some or all of the apparent near-surface, outwardly decreasing, rotation gradient -- but *not* the overall radial or angular differential rotation in the interior, $r < 0.95$, of the convection zone.

The photonic angular momentum loss rate evidently corresponds to a torque that acts only at the photosphere over about a density (and radiation) scale height. The photonic tangential momentum flux at the Sun's surface, $P_d$ then satisfies

$$P_d = \frac{F}{c}\left(\frac{v}{c}\right) \qquad (3)$$

where $F$ is the outward radiative energy flux, $c$ is the speed of light, and $v$ is the local toroidal velocity. Integrating the corresponding angular momentum flux over the solar surface yields a total angular momentum loss rate, or torque, of

$$\frac{dL}{dt} = 2P\Omega R^2/3c^2 \qquad (4)$$

where $\Omega$ is the solar mean surface angular rotation rate and $P$ is the Sun's luminosity. It is interesting that the radiated angular momentum from Eq. (3) has the same form as the Poynting-Robertson drag on, for example, orbiting solar system dust that scatters solar radiation [22,23]. The radiation drag is a relativistic effect [22] but it appears as a photon momentum anisotropy in the Sun's rest frame. For example, we would observe that the east limb of the Sun is slightly bluer and $\Delta T = Tv/c = 0.08$K hotter than the west limb (here $T \approx 5700$K is the photosphere's temperature).

A photospheric torque must be supported by the viscous shear stress (the effect of magnetic fields will be considered elsewhere). In spherical geometry we relate the radial velocity gradient to the shear stress, $\tau_{\phi r}$, and effective viscosity with

$P_d = \tau_{\phi r} = -\frac{\mu \partial v(r)}{\partial r} + \frac{\mu v(r)}{r}$, where μ is the plasma dynamic viscosity and $v(r)$ is the rotation velocity. In this case the second term in the sum on the right is more than an order of magnitude smaller and can be neglected. The photospheric drag and the consequent velocity shear nominally occur over a density scale height where the hydrogen viscosity implies a Reynolds number much larger than unity and therefore turbulent conditions. Nevertheless, equating $P_d$ and $\tau_{\varphi r}$ yields a lower limit for the viscosity. Using solar quantities and the measured velocity gradient from Table 1, we obtain $\mu_p = -F/c^2 \frac{\partial \ln v}{\partial r} \approx 10^{-3}$ kg/m-s. For reference, at photospheric temperatures, the molecular viscosity of hydrogen is about $5 \times 10^{-5}$ (SI), somewhat less than $\mu_p$. In general, we expect turbulent eddies to create a viscosity of order $\mu \sim \rho v l$ where $v$ and $l$ are the characteristic eddy velocity and length scales. Reasonable estimates for these quantities in the Sun's convection zone near the radiative photosphere are $\rho \approx 10^{-4}$, $v \approx 100$, and $l \approx 10^5$ giving $\mu \approx 10^3$ (SI units). Thus, below the photosphere the turbulent viscosity is much larger than typical molecular values and increases rapidly because of the density and scale length. In this simple picture, deeper shear velocity gradients would be smaller than in the photosphere, as is observed.

The total torque calculated from Eq. (4) is $3.6 \times 10^{21}$ (SI units), which is several orders of magnitude smaller than the angular momentum loss rate due to, for example, the solar wind. On the other hand, this drag couples only to the photosphere while the solar wind torque couples through the solar magnetic field at the Alven radius many solar radii out into the corona [24,25]. Thus, it is difficult to see how the solar wind could cause the localized shear we see at the photosphere.

The photosphere's angular momentum would be lost in only a few years if there was no coupling to the convection zone below. Could this weak radiated photon angular momentum loss account for the rotational near-surface slow-down in the outer 5% of the Sun? A linear fit in radius to the interior rotation rate below 0.95 $R_{Sun}$ extrapolated to the surface yields a "baseline" rate for estimating the angular momentum deficit (Fig. 5). The missing angular rotation in the outer 5% corresponds to an angular momentum deficit of $\Delta L = 10^{38}$ (SI units). Surprisingly, this weak radiative damping torque (Eq. 4) times the $5 \times 10^9$ year lifetime of the Sun is a few times larger than $\Delta L$. The local angular momentum loss toward the poles in the Sun also scales like the surface velocity dependence of the rotation damping torque implied by Eq. 3. Given the dynamical changes in the Sun over its lifetime this order of magnitude agreement is, perhaps, surprising but merits further investigation.

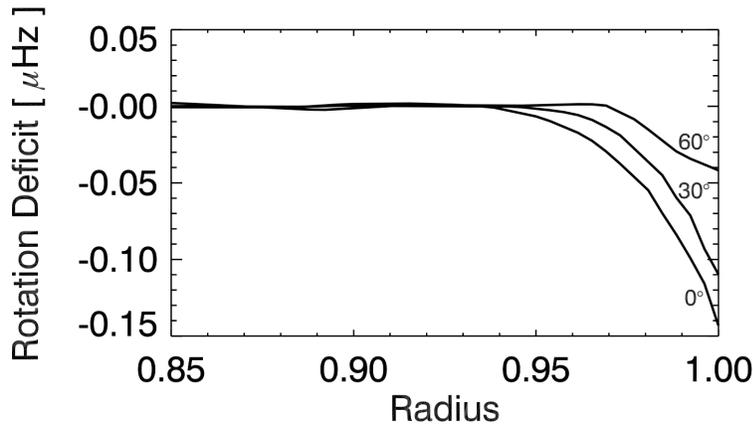

Figure 5: The rotation deficit (in µHz) from a linear trend in radius in the near-surface shear zone is plotted based on Fig. 4 data for solar latitudes 0, 30, and 60 degrees.

It is also interesting to speculate on the radial form of such a rotation deficit. For example, a surface-damped viscous rotating sphere satisfies an *incompressible* Navier-Stokes equation of the form $\frac{d^2(rv)}{rdr^2} - \frac{v}{r^2 sin^2(\theta)} = 0$. This follows from Eq. 15.18c in ref. 26 with steady toroidal flow.

Describing compressible convection with a parametrized turbulent viscosity is a severe approximation, but for a thin outer layer of the Sun it seems to provides qualitative insight. In this case, the solution far from the poles (where the second term is ignorable) is simply $v(r) = constant$ or $\frac{d\log\Omega}{d\log r} = -1$. Also, near the poles the second term becomes important and solutions for $\alpha > -1$ exist, as observed [5]. Thus, photon braking at the surface and turbulent viscous coupling to the interior seem not inconsistent with the radial form of the helioseismic near-surface shear measurements.

**Acknowledgments**

We are grateful to the Stanford and NASA Joint Science Operations Center (JSOC) and the Stanford HMI science team for their support to acquire and understand the HMI data, and for their broad support of the limb science data pipeline. This research was funded by NASA through grant no. NNX14AE08G to JRK. Marcelo Emilio was partially supported the NASA grant and by Brazilian Fundação Araucaria grant 228/2014 and CNPq grant 574004/2008-4. Drs. Charlie Lindsey, Phil Scherrer and Svetlana Berdyugina contributed useful comments on the manuscript.

**References and Notes:**

1. J. Christensen-Dalsgaard, J. Schou, ed. E. J. Rolfe, Seismology of the Sun and sun-like stars. *ESA SP* **286**, 149 (1988).
2. T. M. Brown, J. Christensen-Dalsgaard, W. A. Dziembowski, P. Goode, D. O. Gough, C. A. Morrow, Inferring the Sun's internal angular velocity from observed p-mode frequency splittings. *Astrophys. J.* **343**, 526-546 (1989).
3. J. Schou, H. M. Antia, S. Basu, R. S. Bogart, R. I. Bush, S. M. Chitre, J. Christensen-Dalsgaard, M. P. Di Mauro, W. A. Dziembowski, A. Eff-Darwich, D. O. Gough, D. A. Haber, J. T. Hoeksema, R. Howe, S. G. Korzennik, A. G. Kosovichev, R. M. Larsen, F. P. Pijpers, P. H. Scherrer, T. Sekii, T. D. Tarbell, A. M. Title, M. J. Thompson, J. Toomre, Helioseismic studies of differential rotation in the solar envelope by the solar oscillations investigation using the Michelson Doppler Imager. *Astrophys. J.* **505**, 390-417 (1998).
4. M. J. Thompson, J. Toomre, E. R. Anderson, H. M. Antia, G. Berthomieu, D. Burtonclay, S. M. Chitre, J. Christensen-Dalsgaard, T. Corbard, M. De Rosa, C. R. Genovese, D. O. Gough, D. A. Haber, J. W. Harvey, F. Hill, R. Howe, S. G. Korzennik, A. G. Kosovichev, J. W. Leibacher, F. P. Pijpers, J. Provost, E. J. Rhodes Jr., J. Schou, T. Sekii, P. B. Stark, P. R. Wilson, Differential rotation and dynamics of the solar interior. *Science* **272**, 1300-1305 (1996).
5. T. Corbard, M. J. Thompson, M. J., The subsurface radial gradient of solar angular velocity from MDI f-mode observations. *Sol. Phys.* **205**, 211-229 (2002).
6. A. Barekat, J. Schou, L. Gizon, The radial gradient of the near-surface shear layer of the Sun. *Astron. Astrophys.* **570**, 4 (2014).

7. G. Rüdiger, L. L. Kitchatinov, M. Küker, The surface rotation-laws of young solar-type stars. *Stellar Clusters and Associations: Convection, Rotation, and Dynamos. Proceedings from ASP Conference* **198**, 365 (2000).
8. L. L. Kitchatinov, The solar dynamo: Inferences from observations and modeling. *Geomagnetism and Aeronomy* **54**, 867-876 (2014).
9. M. S. Miesch, B. W. Hindman, Gyroscopic pumping in the solar near-surface shear layer. *Astrophys. J.* **743**, 25 (2011).
10. W. Livingston, R. Milkey, Solar rotation: The photospheric height gradient. *Sol. Phys.* **25**, 267-273 (1972).
11. H. B. Snodgrass, Synoptic observations of large scale velocity patterns on the Sun. *The solar cycle; Proceedings of the National Solar Observatory/Sacramento Peak 12th Summer Workshop, ASP Conference Series* **25**, 205 (1992).
12. J. R. Kuhn, R. Bush, M. Emilio, I. F. Scholl, The precise solar shape and its variability. *Science* **337**, 1638 (2012).
13. J. Schou.; Scherrer, P. H.; Bush, R. I.; Wachter, R.; Couvidat, S.; Rabello-Soares, M. C.; Bogart, R. S.; Hoeksema, J. T.; Liu, Y.; Duvall, T. L.; Akin, D. J.; Allard, B. A.; Miles, J. W.; Rearden, R.; Shine, R. A.; Tarbell, T. D.; Title, A. M.; Wolfson, C. J.; Elmore, D. F.; Norton, A. A.; Tomczyk, S.; Design and Ground Calibration of the Helioseismic and Magnetic Imager (HMI) Instrument on the Solar Dynamics Observatory (SDO), *Sol. Phys.*, **275**, 229-259 (2012).
14. M. Emilio; Couvidat, S.; Bush, R.I.; Kuhn, J. R.; Scholl, I.F; Measuring the Solar Radius from Space During the 2012 Venus Transit, *Astrophys. J.* **798**, 48-55 (2015).
15. A. A. Norton, J. P. Graham, R. K. Ulrich, J. Schou, S. Tomczyk, Y. Liu, B. W. Lites, A. López Ariste, R. I. Bush, H. Socas-Navarro, P. H. Scherrer, Spectral line selection for HMI: A comparison of Fe I 6173 Å and Ni I 6768 Å. *Sol. Phys.* **239**, 69-91 (2006).
16. B. Fleck, S. Couvidat, T. Straus, On the formation height of the SDO/HMI Fe 6173 Å doppler signal. *Sol. Phys.* **271**, 27-40 (2011).
17. J. Christensen-Dalsgaard, Helioseismology, *Rev. Mod. Phys.* **74**, 1073-1129 (2002)
18. S. G. Korzennik, A mode-fitting methodology optimized for very long helioseismic time series. *Astrophys. J.* **626**, 585-615 (2005).
19. R. P. Kraft, Studies of stellar rotation. V. The dependence of rotation on age among aolar-type stars. *Astrophys. J.* 150, 551-570 (1967).
20. G. Rüdiger, M. Kuker, I. Tereshin. The existence of the Lambda effect in the solar convection zone as indicated by SDO/HMI data. *Astron. Astrophys.* 572, L7-11 (2014).
21. S.B. Brun, M. Miesch, J. Toomre, Modeling the dynamical coupling of solar convection with the radiative interior, *Astrophys. J.*, **742**, 79-98 (2011).
22. H. P. Robertson, Dynamical effects of radiation in the solar system. *Mon. Not. R. Astron. Soc.* **97**, 423 (1937).
23. J. A. Burns, P. L. Lamy, S. Soter, Radiation forces on small particles in the solar system. *Icarus* **40**, 1-48 (1979).
24. R. H. Dicke, The Sun's rotation and relativity. *Nature*, **202**, 432-435 (1964).
25. E. J. Weber and L. Davis, The Angular Momentum of the Solar Wind, *Astrophys, J.*, 148, 217-227 (1965).
26. L.D. Landau and E. M. Lifshitz, Fluid Mechanics, Pergamon Press, Oxford, Section 15, (1959)

**Author Contributions:** Ian Cunnyngham conducted most of the analysis of the limb data pipeline output, Dr. Marcelo Emilio analyzed 2D limb spectra for full-disk comparisons, Dr. Isabelle Scholl oversaw the development of the limb data pipeline and database, Dr. Rock Bush provided support from within the Stanford HMI project for a better understanding of the instrument and data, and Dr. Jeff Kuhn directed and was involved in most phases of this program, and bears primary responsibility for errors or omissions in the analysis and results.

**Author Information:** Correspondence and requests for materials should be addressed to J.R.K. (jeff.reykuhn@yahoo.com)

**Table 1. Photospheric angular rotation rates and effective heights derived from limb p-modes**

| Mean height [km] | Rotation rate [µHz] |
|---|---|
| 20 | $2.403 \pm .005$ |
| 80 | $2.399 \pm .001$ |
| 156 | $2.339 \pm .008$ |